\documentclass [12pt,a4paper,draft]{article}
\usepackage{times}

\DeclareFontFamily{OT1}{times}{}
\DeclareFontShape {OT1}{times}{m }{n }{ <-> ptmr }{}
\DeclareFontShape {OT1}{times}{bx}{n }{ <-> ptmb }{}
\DeclareFontShape {OT1}{times}{m }{it}{ <-> ptmri}{}
\DeclareFontShape {OT1}{times}{bx}{it}{ <-> ptmbi}{}
\usepackage{amsfonts}
\newcommand{\cl}{C \kern -0.1em \ell}
\begin{document}

{\LARGE {\bf \noindent Comment on Formulating and Generalizing Dirac's, Proca's, and Maxwell's Equations with Biquaternions or Clifford Numbers}\footnote{\bf Dedicated to Freeman Dyson.}}

\vspace{1\baselineskip}

{\bf \noindent Andr\'e Gsponer and Jean-Pierre Hurni}\footnote{Independent Scientific Research Institute, Box 30, CH-1211 Geneva-12, Switzerland}\\e-mail: isri@vtx.ch

\vspace{1\baselineskip}
{\it Published in} ~~~ {\bf Foundations of Physics Letters, Vol.~14, No.~1 (2001) 77--85.}
\vspace{1\baselineskip}

{\small \noindent \emph{Many difficulties of interpretation met by contemporary researchers attempting to recast or generalize Dirac's, Proca's, or Maxwell's theories using biquaternions or Clifford numbers have been encountered long ago by a number of physicists including Lanczos, Proca, and Einstein. In the modern approach initiated by G\"ursey, these difficulties are solved by recognizing that most generalizations lead to theories describing superpositions of particles of different intrinsic spin and isospin, so that the correct interpretation emerges from the requirement of full Poincar\'e covariance, including space and time reversal, as well as of reversion and gauge invariance.  For instance, the doubling of the number of solutions implied by the simplest generalization of Dirac's equation (i.e., Lanczos's equation) can be interpreted as \emph{isospin}.  In this approach, biquaternions and Clifford numbers become powerful opportunities to formulate the Standard Model of elementary particles, as well as many of its possible generalizations, in very elegant and compact ways.}}

\vspace{2\baselineskip}
{\bf \noindent 1. INTRODUCTION}
\vspace{1\baselineskip}

In the years that followed its discovery, various attempts were made to recast Dirac's equation (1928) using biquaternions (Lanczos, 1929), spinors (van der Waerden, 1929), Clifford numbers (Proca, 1930), semivectors (Einstein, 1932), etc.$^{(1,2)}$  However, only the spinor and the original matrix formulations gained a wide acceptance.

In the 1950-1960's, biquaternion and Clifford number formulations were revived by G\"ursey, Riesz, and Hestenes.  At the same time, the Dirac matrix formulation was more and more often taught with reference to the $\cl_{4,1}$ Clifford algebra,  and  Feynman made some effort to introduce the two-component formalism based on the Pauli algebra  $M_2(\mathbb{C}) \sim \cl_{3,0}$\,.

Yet it is only in last two decades of the XXth century that a growing number of researchers became truly interested in the formulation of fundamental physics using Clifford numbers. While  such formulations have several advantages (e.g., they generally lead to neat expressions that are often very suggestive), they are affected by a number of difficulties, and they often lead to speculative interpretations, which tend to hinder their acceptance by the research community at large.

 In particular, compared to ordinary matrix or tensor formulations, there are: \emph{(i)} losses of information because Clifford numbers do not carry spinor or tensor indices to indicate their behavior under coordinate transformations, (ii) ambiguities due to the noncommutative nature of Clifford numbers (reversion\footnote{\emph{Reversion} is also called \emph{ordinal conjugation} in the quaternion literature because it consists of reversing the order of the factors in a monomial.}) and to the residual arbitrariness in their definition,$^{(2,3)}$ \emph{(iii)} difficulties of interpretation coming from the increase in the number of components when vectors or spinors are  replaced by Clifford numbers,$^{(4,5)}$ \emph{(iv)} similar difficulties when ordinary scalars such as the mass are promoted to Clifford numbers, and \emph{(v)} more fundamental difficulties with the quantum interpretation when real Clifford algebras are used instead of complexified ones. The last difficulty comes from the well known fact that if \emph{``i''} is not explicit in a quantum theory, it must be replaced by the definition of special rules$^{(6)}$ or by an equivalent antiunitary operator.$^{(7)}$

In a short comment it is not possible to address all these difficulties in detail and to refer to the numerous papers in which they arise. (For a selection,  see Ref.8--12 and references therein.)  We therefore take a simple but non trivial case:  the problem of formulating physical theories with biquaternions. Since these are just complexified quaternions, $\mathbb{B} = \mathbb{C} \otimes \mathbb{H} \sim  \cl_{3,0}$\,, there are no difficulties with the quantum interpretation. Nevertheless,  biquaternions are suitable to show how the other difficulties can be circumvented, and how much the isomorphisms $SU(2) \sim \mathbb{H} / \mathbb{R}$ and $SL(2,\mathbb{C}) \sim \mathbb{B}^\star{\! /} \Bbb{C}$ are useful in making explicit the basic symmetries of the Standard Model.

\vspace{2\baselineskip}
{\bf \noindent 2. DIRAC'S AND LANCZOS'S FUNDAMENTAL EQUATIONS}
\vspace{1\baselineskip}

In the standard spinor or two-component formalisms,  Dirac's  equation can be symbolically written as  
$$ 
     \overline\partial  L = m R  \hskip 1 in     \partial  R = m L   \eqno(1) 
$$
where $L$ and $R$ are the left- and right-handed components of Dirac's four-component bispinor, and $\partial$ the spinor or Pauli-matrix 4-gradient.  

The simplest generalization of this equation (which amounts  to replace the two-component spinors $L$ and $R$ by two biquaternions $A$ and $B$) has first been made by Lanczos:
$$
  \overline\nabla A = m B  \hskip 1 in      \nabla B = m A ~~.      \eqno(2)
$$
Here $\nabla = [\partial_{it},\partial_{\vec x}]$  is the quaternion 4-gradient operator,  assumed to  transform  as  a Lorentz 4-vector, i.e., $\nabla^\prime=\mathcal{L}\nabla\mathcal{L}^\dag$ with $\mathcal{L} \in SL(2,\mathbb{C})$. This is \emph{Lanczos's fundamental equation}$^{(4)}$ which has the remarkable property of describing isospin doublets of spin $\frac{1}{2}$ particles, as well as  spin 0 and spin 1 particles of either positive or negative space-reversal parity.  The existence of isospin doublets is the consequence of the invariance of (2) under the $SU(2)$ gauge transformation $ A \rightarrow AG$, $ B \rightarrow BG$.

 Dirac's equation is obtained from (2) by making the superpositions:$^{(4,13)}$
$$
     D_+ = A \sigma + B^* \overline\sigma  \hskip 1 in      D_- = (A \overline\sigma - B^* \sigma) i\vec e_1 \eqno(3)
$$
where $\sigma$ is a constant idempotent.\footnote{For  definitiveness,  we  take  $\sigma=\frac{1}{2}(1+i\vec e_3)$  where $\vec e_3$  is  the  third  quaternion unit, and $\vec e_1$ for a typical unit vector perpendicular to $\vec e_3$.} Both of these linearly independent superpositions satisfy the Klein-Gordon equation,    $\nabla\overline\nabla D = m^2 D$, as well as the  {\it Dirac-Lanczos equation}:  
$$
    \overline\nabla  D = m D^* i \vec e_3 ~~.   \eqno(4)
$$
   This equation is strictly equivalent to Dirac's. It was  first discovered by Lancz\-os$^{(4,~p.462)}$ and later  rediscovered by a number of people,  in particular by G\"ursey$^{(14)}$ and Hestenes.$^{(15)}$  In fact, in a form or another,  it is the basic equation  in any Clifford type formulation of Dirac's theory.

The spin $\frac{1}{2}$ character of the superpositions (3) is enforced by the postmultiplication by $\sigma$, and in (4) by the factor $ie_3$, which imply that under a Lorentz transformation we have $D^\prime=\mathcal{L}D$. 

     The correct interpretation of $D_+$ and $D_-$ as an isospin doublet  has  been  given for the first time by G\"ursey.$^{(13)}$  This interpretation solves the ``doubling'' problem that puzzled Lanczos$^{(4,2)}$ and others more recently.$^{(5,9)}$

\vspace{2\baselineskip}
{\bf \noindent 3. PROCA'S EQUATION AND REVERSION}
\vspace{1\baselineskip}

In order to obtain the spin 0 and spin 1 solutions of (2) one has to address a problem that is implicitly solved by making the superpositions (3) in the spin~$\frac{1}{2}$ case. In effect, a fundamental concept used in expressing a physical theory in mathematical form is that any arbitrariness in the writing of the equations  should have no consequences on the physics. While this principle is automatically taken care of in tensor calculus, in Clifford (or biquaternion) formulations it implies that the \emph{order} of the factors (from left to right, or right to left) should have no consequencies.$^{(2)}$ In our case, the \emph{reversed} version of equation (2)
$$
  A^\sim \overline\nabla = m B^\sim  \hskip 1 in     B^\sim \nabla  = m A^\sim \eqno(5)
$$
could equally well have been taken as \emph{Lanczos's fundamental equation}. As shown in Ref.2, the concepts of reversion and space reversal are intimately connected: A requirement of definite reversion symmetry such as $A^\sim = \pm A$ corresponds to selecting fields of even or odd parity. For  example,  if $A=A^\sim$ is an odd parity 4-vector, it may represent Maxwell's or Proca's 4-potential. In this case, the corresponding reversion invariant field equation is obtained by combining (2) and (5):
$$
 \overline\nabla A = m B  \hskip 1 in    \nabla B + B^\sim \nabla = 2 m A  ~~. \eqno(6)
$$
  However, while the fields in (6) have a well defined parity, they are  still mixtures of spin 0 and spin 1 states. This is because under a Lorentz transformation  we have  $B^\prime=\mathcal{L}^\ast B\mathcal{L}^\dag$, so that the scalar part of $B$ is a Lorentz  invariant,  while the vector part transforms as a Lorentz 6-vector.  Hence, the field  $[b,\vec b\,] = mB$  can be covariantly decomposed into its scalar and vector  parts.   Introducing the  corresponding  decomposition  of  the  potential $A=A_0+A_1$, the system (6) can be split into a spin 0 and  a spin 1 part.

     For the {\it spin zero} part, because $A=A^\sim=A_0$, we see from (2,5)  that $b=b^\sim$.  The system (6) reduces therefore to  the  following   degenerate form of Lanczos's equation: 
$$
  \overline\nabla \cdot A_0 = b   \hskip 1 in   \nabla b = m^2 A_0 ~~. \eqno(7)
$$

      For the {\it spin one} part,   the resulting equation is just (6)  with  the  field  $B$ restricted to its vector  part $\vec b$.   This  is  {\it Proca's equation} for a massive spin 1 particle:$^{(2)}$ 
$$
 \overline\nabla \wedge A_1 = \vec b  \hskip 1 in \nabla\vec b  + \vec b^\sim\nabla = 2 m^2 A_1 ~~. \eqno(8)
$$

     The   key  assumption  in  going  from  Lanczos's  fundamental  equation (2,5) to (7,8),  i.e.,  the field equations for  a  massive  ``scalar'' or ``vector'' particle,  was that $A$ had to be a reversion  symmetric  4-vector.   If we had assumed the second  fully  covariant possibility,  i.e., that $A$ was reversion anti-symmetric, we  would  have  obtained the  field equations for  a  {\it pseudoscalar}  and  respectively  a {\it pseudovector} particle.   The only difference  is  that  in  the  second equation of (8) the plus sign  has  to  be  replaced  by  a  minus sign. The fact that  all  equations  for  particles  of spin 0 and 1 are degenerate cases of  Lanczos's  equation  has first been shown by G\"ursey in his PhD  thesis.$^{(14,~p.162)}$

From Proca's equation (8) it is now straightforward to derive the conserved current and the energy-momentum tensor$^{(2)}$
$$
    C = {A_1}^\dag \vec b + {\vec b}^\dag  A_1 + (...)^\sim \eqno(9)
$$
$$
  8\pi T(\;) = \vec b^\dag  (\;)\vec b + m^2 A_1^\dag (\;)A_1 + (...)^\sim \eqno(10)
$$
where  $(...)^\sim$ means that  the expression  has to be completed  by adding the reversed version of the part on the left.  If now the electromagnetic field is introduced, either by local  gauge  invariance  or the  minimum  coupling substitution ansatz,  the divergence of $C$ will still be zero, while the divergence of $T(\;)$ will give the Lorentz force acting on the particle.

\vspace{2\baselineskip}
{\bf \noindent 4. MAXWELL'S EQUATIONS AND REVERSION}
\vspace{1\baselineskip}

  Now that we have derived Proca's  equation,  we  can set $m=0$ in   (8)  to get the quaternionic  field  equation of a massless particle of spin~1.   Including a  source  term with a current $J$ we obtain: 
$$
 \overline\nabla \wedge \phi = \vec f  \hskip 1 in \nabla\vec f  + \vec f^\sim\nabla = -8\pi J ~~. \eqno(11)
$$
 This is the correct form of {\it Maxwell's second set of equations} in quaternion form. It   explicitly  implies  that  the  source  current $J$,  and  thus  the  potential $\phi$,  has  to  be  reversion  symmetric  and  shows  that  the  electromagnetic  field  is the antisymmetric  tensor:
$$
       2 F(\;) = (\;)\vec f + \vec f^\sim(\;)  ~~.  \eqno(12)
$$
 Contrary to the 6-vector $\vec f = \vec E + i \vec B$,  this tensor is reversion symmetric, and  therefore defines a physical field of odd parity.   Its correct  quaternionic form was first given by Kilmister.$^{(3)}$

     The  usual quaternion form of Maxwell's equations contains  the  electromagnetic  6-vector $\vec f$ instead of the electromagnetic tensor  (12): 
$$
 \overline\nabla \wedge \phi = \vec f  \hskip 1 in \nabla\vec f = -4\pi J ~~. \eqno(13)
$$
 This  equation is formally incorrect because it is not  covariant  under  spatial  reversal.   However, provided that $\phi$ and $J$ are postulated  to be reversion invariant,  it  is  in  practice equivalent to (11) because from Lanczos's fundamental system (2,5) we have then:
$$
      \nabla \vec f - \vec f^\sim \nabla = 0 ~~. \eqno(14)
$$
 This equation corresponds to the requirement that there should be no magnetic  pole or current, i.e., to {\it Maxwell's first set of equations}. Equation (14), however, is stronger than these equations in tensor form: it totally excludes any kind of magnetic source.  Therefore, Lanczos's equation and reversion invariance enforce a fundamental asymmetry that is strongly supported by experiment, i.e., the existence of point-like electric charges, and the absence of magnetic monopoles.$^{(2)}$

\vspace{2\baselineskip}
{\bf \noindent 5. EINSTEIN-MAYER'S EQUATION AND STANDARD MODEL}
\vspace{1\baselineskip}

Looking at (2), an obvious generalization is to assume that the mass $m$ is a quaternionic parameter or field. This assumption leads to the concept of \emph{hypercomplex mass},$^{(5)}$ or to the \emph{versatile Dirac equation},$^{(8)}$ and was first made by Einstein and Mayer when they generalized Lanczos's equation as follows:$^{(16)}$ 
$$
   \overline\nabla A  = B E^\dag  \hskip 1 in  \nabla B  = A E ~~. \eqno(15)
$$
Depending on the suppositions made for $E\in \mathbb{B}$ one has a rich spectrum of solutions. As with Lanczos's fundamental system, these have to be classified according to spin, parity, and mass. In field theory terminology, when $E$ is a spin~0 field such as the Higgs doublet of the Standard Model, (15) corresponds to a ``Yukawa coupling'' between $E$ and the $A$ and $B$ fields.$^{(17)}$

     A  major difference with Lanczos's fundamental system is  that  the second order equations for $A$ and $B$ are  now  eigenvalue  equations for the mass. In other words, while all solutions of (2) have the same mass, (15) has pairs of solutions with different masses that Einstein and Mayer tried to interpret as electron-proton doublets.$^{(16)}$ In the modern perspective, two particularly interesting special cases emerge: First, when $EE^\dag=m^2$ the spin $\frac{1}{2}$ isospin-doublet can be interpreted as a proton-neutron pair (because it is degenerate in mass). Then, if $E$ is assumed to be a pseudoscalar field, one is lead to the charge independent theory of low-energy nuclear forces.$^{(18)}$  Second, when $E\overline{E}=0$ the doublet can be interpreted as an electron-neutrino pair (because one of the two masses is necessarily zero).  Then, assuming a parity violating  $SU(2)$ local gauge  such  that $A \to AG$,  $B \to B$ and $E \to G^\dag E$, one gets a low-energy weak interaction phenomenology that is equivalent to  that  of  the  Standard Model.$^{(19)}$

 Other physical interpretations are possible when $E$ is derivatively coupled to the $A$ and $B$ fields. For example, if
$$
                \nabla E = A \overline B              \eqno(16)
$$
the system (15-16) leads to double-periodic Petiau waves. These solutions  interpolate  between  pure de Broglie waves  and  pure  solitonic  waves:  a  beautiful  realization  of  the  wave/particle duality of quantum mechanics.$^{(20)}$

However, possibly the most interesting interpretation is to relate $E$ to the Higgs field of the Standard Model:
$$
       \nabla\overline{\nabla} E = - \mu E + \lambda E{E^\dag}E ~~.   \eqno(17)
$$
Since this equation is invariant in the quaternion gauge transformation $ E \rightarrow {G^\dag}EH$ it has the manifest $SU(2)_L \times SU(2)_R$ symmetry of the $\sigma$-Model which, after breaking the $SU(2)_R$ symmetry, leads to the renormalizable electroweak theory of the Standard Model.$^{(21)}$

\vspace{2\baselineskip}
{\bf \noindent 6. CONCLUSION}
\vspace{1\baselineskip}

In this comment we have shown that the use of Clifford algebras, and more particularly of biquaternions, can lead to a satisfactory formulation of elementary particle physics which, in the Standard Model, is entirely based on fundamental fields of spin 0, $\frac{1}{2}$ and 1. In order to build such a formulation on solid ground, it is essential to require that all fields are irreducible representations of the full Poincar\'e group, including the discrete space and time reversal symmetries; as well as covariant under reversion, a requirement that is specific to any theory formulated with Clifford numbers.

 Under these conditions, many of the difficulties encountered when using non-commutative Clifford numbers instead of real or complex numbers can be overcome. Moreover, most (if not all) apparently new results can be interpreted using standard concepts and without introducing ``new'' physics. 

 Finally, we stress again the very interesting fact that the use of the most simple Clifford algebra directly applicable to quantum physics, i.e., the biquaternions, leads in a natural way to the existence of isospin doublets, $SU(2)$ gauge fields, and other key features of the current Standard Model. In this respect, the predictive power of Lanczos's equation and its generalizations by Einstein and others, as well as the many opportunities opened by the use of biquaternions and Clifford numbers in physics, are most remarquable.$^{(22)}$

\vspace{2\baselineskip}
{\bf \noindent REFERENCES} 
\vspace{1\baselineskip}
\begin{enumerate}
%
\setlength{\parskip}{-1.4mm}
\item W.R. Davis et al., eds., \emph{Cornelius Lanczos Collected Published Papers With Commentaries} (North Carolina State University, Raleigh, 1998). Web page {www.physics.ncsu.edu/lanczos}.
\item A. Gsponer and J.-P. Hurni, ``Lanczos-Einstein-Petiau: From Dirac's equation to non-linear wave mechanics,'' in Ref.1, pages 2-1248 to 2-1277.
\item C.W. Kilmister, \emph{Proc. Roy. Irish Acad.}  {\bf 57}, 37--52 (1955).
\item C. Lanczos, \emph{Zeits. f. Phys.} {\bf 57}, 447--473, 474--483, 484--493 (1929). Reprinted and translated in Ref.1, pages 2-1132 to 2-1225.
\item J.D. Edmonds, Jr., \emph{Int. J. Theor. Phys. } {\bf 6}, 205--224 (1972); \emph{Found. Phys. } {\bf 3}, 313--319 (1973).
\item C. Lanczos, \emph{Zeits. f. Phys.} {\bf 37}, 405--513 (1926). Reprinted and translated in Ref.1, pages 2-932 to 2-949.
\item E.C.G. Stueckelberg et al., \emph{Helv. Phys. Acta} {\bf 33}, 727--752 (1960); {\bf 34}, 621--628 (1961).  See also section 2.6 in S.L. Adler, \emph{Quaternionic Quantum Mechanics and Quantum Fields} (Oxford University Press, New York, 1995).
\item S.B.M. Bell, J.P. Cullerne, and B.M. Diaz, \emph{Found. Phys.} {\bf 30}, 35--57 (2000).
\item S. De Leo and W.A. Rodrigues, \emph{Int. J. Theor. Phys.} {\bf 37}, 1511--1529, 1707--1720 (1998).
\item J. Keller and Z. Oziewicz, eds., \emph{ Proc. of the Int. Conf. on the Theory of the Electron}, Adv. in Appl. Clifford Algebras {\bf 7(S)} (1997).
\item S. Gull et al., \emph{Found. Phys.} {\bf 23}, 1175--1201 (1993).
\item P. Lounesto, \emph{Found. Phys.} {\bf 23}, 1203--1237 (1993).
\item F. G\"ursey, \emph{Nuovo Cim.}  {\bf 7}, 411--415 (1958). 
\item F. G\"ursey, PhD thesis, (University of London, 1950).
 See also F. G\"ursey, \emph{Phys. Rev.} {\bf 77}, 844 (1950).  
\item D. Hestenes,  \emph{Space-Time  Algebra} (Gordon and Breach, New York, 1966, 1987, 1992); \emph{J. Math. Phys.} {\bf 8}, 798--808, 809--812 (1967).
\item A. Einstein and W. Mayer, \emph{Sitzber. der Preuss. Akad}, 522--550 (1932);  \emph{Proc. Roy. Acad. Amsterdam} {\bf 36}, 497--516,  615--619 (1933). 
\item See, for example, T.P. Cheng and L.F. Li, \emph{Gauge theory of elementary particle physics} (Clarendon Press, Oxford, 1984).
\item F. G\"ursey, \emph{Nuovo Cim.} {\bf 16}, 230--240 (1960).
\item See,  for example,  J.D. Bjorken,  \emph{Phys. Rev.} {\bf D 19}, 335--346 (1979).
\item G. Petiau,   \emph{Nuovo Cim.} {\bf 40}, 84--101 (1965); \emph{Ann. Inst. Henri Poincar\'e} {\bf 36}, 89--125 (1982).
\item M. Veltman, \emph{Reflexions on the Higgs system}, CERN Report 97-05 (1997).
\item F.J. Dyson, ``Missed opportunities,'' \emph{Bull. Am. Math. Soc.} {\bf 78}, 635--652 (1972); ``The threefold way,'' \emph{J. Math. Phys.} {\bf 6}, 1199--1215 (1962).
\end{enumerate}
\end{document}